\documentclass[pra,twocolumn,superscriptaddress]{revtex4-1}

\usepackage[utf8]{inputenc}
\usepackage[english]{babel}
\usepackage[T1]{fontenc}
\usepackage{lmodern}

\usepackage{graphicx}
\usepackage{amsmath}
\usepackage{amsfonts}
\usepackage{amssymb}

\usepackage{microtype}

\usepackage[breaklinks=true,colorlinks]{hyperref}

\newcommand{\coloneq}{\mathrel{\mathop:}=}
\newcommand{\eqcolon}{=\mathrel{\mathop:}}
\newcommand{\Tr}{\operatorname{Tr}}
\newcommand{\ket}[1]{\left|{#1}\right\rangle}

\newcommand{\ketbra}[2]{\left|{#1}\middle\rangle\middle\langle{#2}\right|}
\newcommand{\proj}[1]{\ketbra{#1}{#1}}
\newcommand{\realp}{\operatorname{Re}}
\newcommand{\dd}{\mathrm{d}}

\newcommand{\kB}{k_\mathrm{B}}
\newcommand{\nenv}{\bar{n}_\mathrm{env}}
\newcommand{\Tenv}{T_\mathrm{env}}
\newcommand{\omegac}{\omega_\mathrm{c}}
\newcommand{\Tc}{T_\mathrm{c}}
\newcommand{\Ta}{T_\mathrm{a}}
\newcommand{\ttr}{t_\mathrm{tr}}

\begin{document}

\title{Temperature control in dissipative cavities by entangled dimers}

\author{Ceren B. Da\u{g}}
\thanks{These authors contributed equally to this work.}
\affiliation{Physics Department, University of Michigan, Ann Arbor, Michigan 48109, USA}

\author{Wolfgang Niedenzu}
\thanks{These authors contributed equally to this work.}
\email{Wolfgang.Niedenzu@uibk.ac.at}
\affiliation{Institut f\"ur Theoretische Physik, Universit\"at Innsbruck, Technikerstra{\ss}e~21a, A-6020~Innsbruck, Austria}
\affiliation{Department of Chemical Physics, Weizmann Institute of Science, Rehovot 7610001, Israel}

\author{Fatih Ozaydin} 
\affiliation{Department of Information Technologies, Is{\i}k University, 34980 Sile, Istanbul, Turkey}
\affiliation{Photon Science Center, Graduate School of Engineering, The University of Tokyo, 7-3-1 Bunkyo-ku, Tokyo 113-8656, Japan}

\author{\"{O}zg\"{u}r E. M\"{u}stecapl{\i}o\u{g}lu}
\email{omustecap@ku.edu.tr}
\affiliation{Department of Physics, Ko\c{c} University, 34450 Sar{\i}yer, Istanbul, Turkey}
  
\author{Gershon Kurizki} 
\affiliation{Department of Chemical Physics, Weizmann Institute of Science, Rehovot 7610001, Israel}

\date{January 30, 2019}

\begin{abstract}
We show that the temperature of a cavity field can be drastically varied by its interaction with suitably-entangled atom pairs (dimers) traversing the cavity under realistic atomic decoherence. To this end we resort to the hitherto untapped resource of naturally entangled dimers whose state can be simply controlled via molecular dissociation, collisions forming the dimer, or unstable dimers such as positronium. Depending on the chosen state of the dimer, the cavity-field mode can be driven to a steady-state temperature that is either much lower or much higher than the ambient temperature, despite adverse effects of cavity loss and atomic decoherence. Entangled dimers enable much broader range of cavity temperature control than single ``phaseonium'' atoms with coherently-superposed levels. Such dimers are shown to constitute highly caloric fuel that can ensure high efficiency or power in photonic thermal engines. Alternatively, they can serve as controllable thermal baths for quantum simulation of energy exchange in photosynthesis or quantum annealing.
\end{abstract}

\maketitle

\section{Introduction}

Temperature control of reservoirs (baths) that serve as energy sources or entropy dumps is a key tool for understanding the conceptual subtleties of thermodynamics in the quantum realm~\cite{kosloff2013quantum,gelbwaser2015thermodynamics,goold2016role,vinjanampathy2016quantum} and exploiting heat in practical quantum technologies like quantum heat machines~\cite{alicki1979quantum,kosloff1984quantum,geva1996quantum,scully2003extracting,quan2006quantum,liao2010single,abah2012single,li2014quantum,gelbwaser2015thermodynamics,kato2016quantum,rossnagel2016single,vinjanampathy2016quantum,klatzow2017experimental,kosloff2017quantum}, quantum simulators of dissipative systems, such as light harvesting complexes~\cite{huelga2013vibrations,mostame2012quantum,dorfman2013photosynthetic,killoran2015enhancing,wertnik2018optimizing}, or thermal quantum annealers~\cite{boixo2013experimental,boixo2014evidence,shabani2016artificial}. It is particularly intriguing that quantum coherence can be used as a temperature knob of the bath~\cite{scully2003extracting,dillenschneider2009energetics,huang2012effects,abah2014efficiency,rossnagel2014nanoscale,hardal2015superradiant,dag2016multiatom,niedenzu2016operation,turkpence2016quantum,niedenzu2018quantum,klaers2017squeezed}; in addition to its broad variety of applications, the quantum coherent route to engineering artificial thermal baths can offer fundamental insights into the quantum-to-classical transition and the role of ``quantumness'' in energy exchange and computation processes~\cite{zurek2003decoherence}.

\par

Here we investigate the temperature control of a simple dissipative quantum system: a leaky cavity field mode that interacts with a quantum-entangled system, which is also subject to decoherence, for simplicity an ensemble of correlated pairs of atoms (dimers). We ask: Does their two-atom coherence (the operational equivalent of entanglement~\cite{streltsov2015measuring,winter2016operational}) affect the heat exchange of the two open systems? And if so, are there benefits to such heat exchange under realistic conditions? The answer to both questions is shown to be positive.

\par

These questions have been investigated for quantum coherence between the lower levels in a three-level atom (``phaseonium'')~\cite{scully2003extracting} and later for atomic clusters~\cite{dag2016multiatom}, but the foregoing investigations have not addressed dissipation or decoherence caused by the environment, so that their experimental relevance is unclear. Furthermore, a systematic exploration of the multipartite coherence effects in an atomic cluster~\cite{dag2016multiatom} is lacking. This prompts us to assess the feasibility and advantages of quantum entangled dimers~\cite{dag2016multiatom} for cavity temperature control as compared to a micromaser pumped by phaseonium atoms or by two-level atoms (TLAs) (one at a time) whose temperature is controllable by the population ratio of the excited to ground-state levels~\cite{scully1967quantum,filipowicz1986theory,scullybook,meystrebook}.

\par

The basis for this comparison is that a TLA in a typical atomic beam can carry more excitation than either a phaseonium atom~\cite{scully2003extracting} or an entangled dimer~\cite{dag2016multiatom} in a micromaser and yet the latter (coherent) resources can yield higher temperatures of the cavity field, owing to their coherence. Since the maximum value of coherence depends on the atomic level populations, the resulting temperature range that can be generated in a leaky cavity by coherent resources, particularly entangled dimers, is a key piece of the puzzle that we address in Sec.~\ref{sec_temperature}. The central result~\eqref{eq_cooling_heating} is the ambient temperature range for which quantum-coherence-to-heat-conversion by atomic dimers is beneficial.

In Sec.~\ref{sec_dimers} we show that the required entanglement between two atoms is controllable in a straightforward manner by dissociation of molecular dimers~\cite{kurizki1985quantum,kurizki1987theory,fry1995proposal,urbanczyk2012entangled} or atom-atom collisions in a cavity~\cite{deb1999formation}. In addition, unstable bound states of electron--positron pairs (positronium)~\cite{mills1975effects,diehl1991temperature} can be a source of entangled dimers to fuel gamma-ray micromasers~\cite{rich1981recent,acin2001three,harpen2003positronium,mills2004prospects,thomas2014optical}. Our comparative examination shows that entangled dimers, upon allowing for cavity leakage and dimer decoherence, are able to provide a remarkably versatile cavity temperature control, covering a broader temperature range than previously proposed single-atom~\cite{scully2003extracting} or two-atom~\cite{dillenschneider2009energetics} coherent quantum fuels (Secs.~\ref{sec_decoherence} and~\ref{sec_comparison_phaseonium}).

\par

The above findings identify entangled dimers as an advantageous quantum fuel that may endow photonic heat engines or refrigerators with very high efficiency. Its other possible applications may include quantum simulators for light harvesting complexes~\cite{huelga2013vibrations,mostame2012quantum,dorfman2013photosynthetic,killoran2015enhancing,wertnik2018optimizing}, or thermal quantum annealers~\cite{boixo2013experimental,boixo2014evidence,shabani2016artificial}. We conclude and discuss our results, together with suggestions of potential applications, in Sec.~\ref{sec_discussion}. The present work could be a stepping stone towards realistic implementations of artificial quantum baths that would give us a far more advanced control over thermal processes at the quantum level than existing setups.

\section{Master equation and cavity temperature}\label{sec_temperature}
\label{sec:theory}

\begin{figure}
  \centering
  \includegraphics[width=0.95\columnwidth]{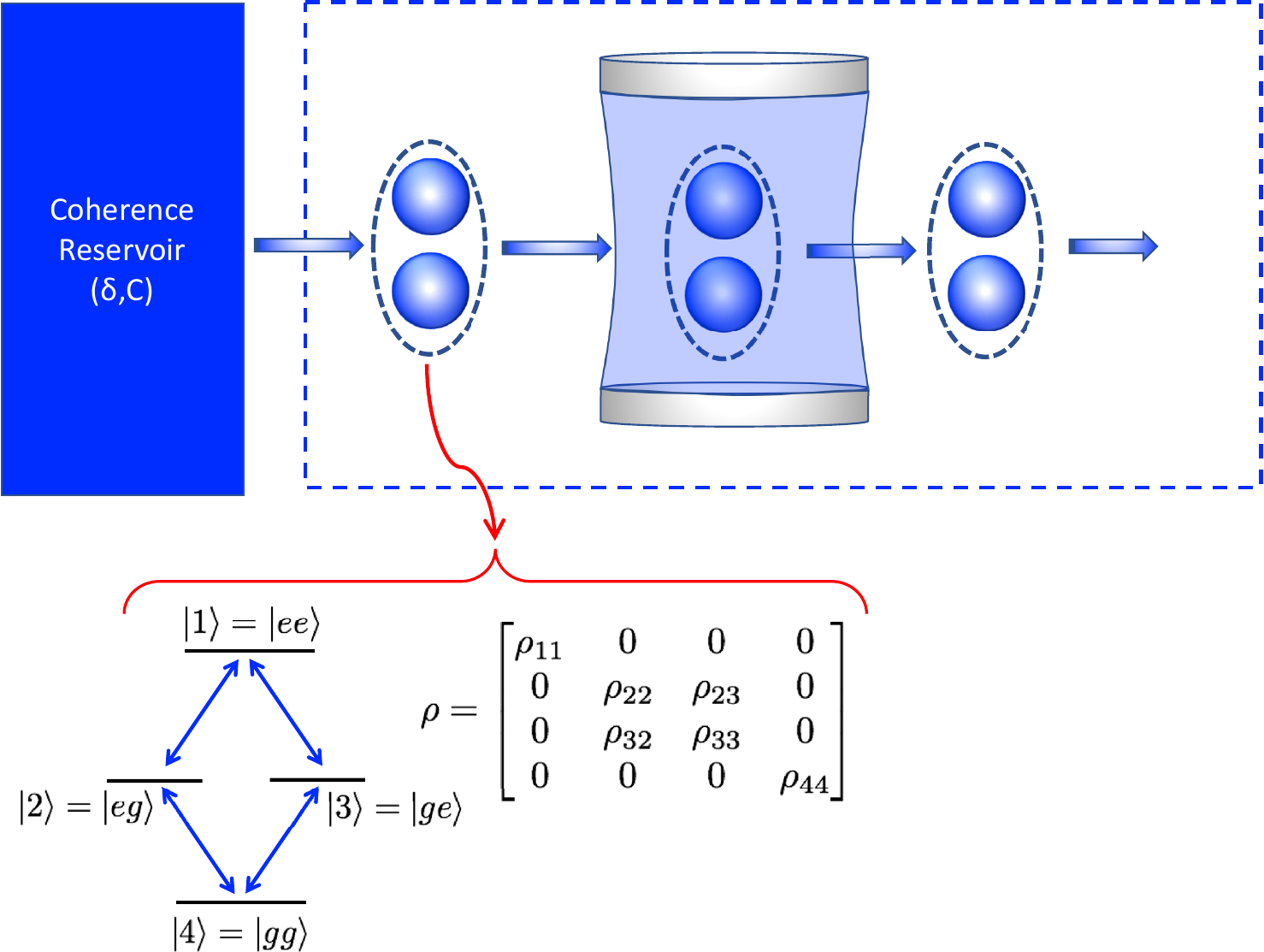}
  \caption{A micromaser powered by a quantum fuel with heat exchange coherences (HECs). The setup consists of a cavity resonantly pumped by a beam constituted of two-level atom (TLA) pairs. The beam drives the cavity field into a thermal state whose temperature $\Tc$ is determined by the atom-pair (double-excitation) population inversion $\delta\coloneq 2(\rho_{11}-\rho_{44})$ and the coherence $C\coloneq 2\realp{\rho_{23}}$.}\label{fig_hec_scheme}
\end{figure}

Let us consider a beam of dimers injected into a cavity (see Fig.~\ref{fig_hec_scheme}), where every dimer is prepared in an initial state of the form
\begin{equation}\label{eq_rho}
  \rho=
  \begin{pmatrix}
    \rho_{11} & 0 & 0 & 0 \\
    0 & \rho_{22} & \rho_{23} & 0 \\
    0 & \rho_{32} & \rho_{33} & 0 \\
    0 & 0 & 0 & \rho_{44}
  \end{pmatrix}
\end{equation}
in the basis of the vectors $\{\ket{1},\ket{2},\ket{3},\ket{4}\}\coloneq\{\ket{ee},\ket{eg},\ket{ge},\ket{gg}\}$, where $\ket{g}$ and $\ket{e}$ are the ground- and excited states of the dimer constituents (two identical TLAs). The state~\eqref{eq_rho} is quantum-entangled if at least one eigenvalue of the partially-transposed $\rho$ is negative~\cite{peres1996separability}, which in turn holds iff $|\rho_{23}|^2>\rho_{11}\rho_{44}$. Conversely, for $|\rho_{23}|^2\leq\rho_{11}\rho_{44}$ the state $\rho$ is only classically correlated and thus separable. The only non-vanishing coherence $\rho_{23}$ has been dubbed by us a ``heat exchange coherence'' (HEC)~\cite{dag2016multiatom}, meaning that it has the potential to change the temperature of the cavity mode. By contrast, any other non-vanishing coherence would contribute either to the mean-amplitude displacement or squeezing of the intracavity field~\cite{dag2016multiatom}.

\par

Under the standard assumptions of micromaser theory~\cite{filipowicz1986theory,scullybook,meystrebook}, the master equation describing the cavity field dynamics for resonant interaction is given by~\cite{dag2016multiatom} (see Appendix~\ref{app:derivation})
\begin{equation}\label{eq_master}
  \dot\rho_\mathrm{c}=\left[\frac{\mu r_-}{2}+\kappa(\nenv+1)\right]\mathcal{L}_\mathrm{d}\rho_\mathrm{c}+\left[\frac{\mu r_+}{2}+\kappa\nenv\right]\mathcal{L}_\mathrm{e}\rho_\mathrm{c}.
\end{equation}
Here $\mathcal{L}_\mathrm{d}\rho_\mathrm{c} = 2 a \rho_\mathrm{c} a^{\dagger} - a^{\dagger} a \rho_\mathrm{c} - \rho_\mathrm{c}  a^{\dagger} a$ and $\mathcal{L}_\mathrm{e}\rho_\mathrm{c} = 2 a^{\dagger} \rho_\mathrm{c} a - a a^{\dagger} \rho_\mathrm{c} - \rho_\mathrm{c} a a^{\dagger}$ are the Lindblad operators for incoherent de-excitation and excitation of the cavity field, respectively, $\kappa$ is the cavity loss rate and $\nenv=\{\exp[\hbar\omega_\mathrm{c}/(\kB \Tenv)]-1\}^{-1}$ is the number of thermal photons in the environment at temperature $\Tenv$. The effect of the atomic beam is encoded in the coefficients
\begin{equation}\label{eq_r_pm}
  r_\pm\coloneq 1+C\pm\frac{\delta}{2}.
\end{equation}
Here $\delta\coloneq 2(\rho_{11}-\rho_{44})$ is the double-excitation population inversion, ranging from $\delta=-2$ for the fully-uninverted state $\rho=\ketbra{gg}{gg}$ to $\delta=2$ for the fully-inverted state $\rho=\ketbra{ee}{ee}$. The double-excitation inversion $\delta$ corresponds to the energy of the dimer in state~\eqref{eq_rho} via $E=\hbar\omegac(1+\delta/2)$ (see Appendix~\ref{app:dimerEnergy}). The two-atom coherence is denoted by $C\coloneq 2\realp\rho_{23}$ (we can w.l.o.g.\ assume that $\rho_{23}$ is real). Finally, $\mu=r(g \tau)^2$ in Eq.~\eqref{eq_master} is an effective coupling rate determined by the injection rate $r$ of atomic pairs, their interaction time $\tau$ with the cavity field and their coupling strength $g$ to the cavity mode. The second-order master equation~\eqref{eq_master} is valid for $\tau\ll 1/g$ and the additional condition $\tau\ll 1/r$ ensures that at most one dimer is present in the cavity at once.

\par

Under the condition of operation below micromaser threshold, $r_+<r_-+2\kappa/\mu$, i.e., $\delta<2\kappa/\mu$, the master equation~\eqref{eq_master} yields a thermal steady state of the cavity field, $\rho_\mathrm{c}^\mathrm{ss}=Z^{-1}\exp[-\hbar\omegac a^\dagger a/(\kB\Tc)]$ with $Z=\Tr\{\exp[-\hbar\omegac a^\dagger a/(\kB\Tc)]\}$, determined by the temperature
\begin{equation}\label{eq_Tc}
  \frac{r_++2\kappa\nenv/\mu}{r_-+2\kappa(\nenv+1)/\mu}\eqcolon \exp\left(-\frac{\hbar\omegac}{\kB \Tc}\right),
\end{equation}
where the l.h.s.\ is the ratio of the absorption and emission coefficients in Eq.~\eqref{eq_master}. Intriguingly, $\Tc$ is a proper temperature for the cavity field (which relaxes to a Gibbs state) although it can be controlled by varying the two-atom coherence $C$ (i.e., the HEC $\rho_{23}$) in the atom-pair beam which is prepared in the highly non-equilibrium state~\eqref{eq_rho}. The beam can nevertheless act as an effective heat bath at temperature $\Tc$ for the cavity field. This temperature is well-defined (positive and finite) only for operation below the micromaser threshold.

\par

\begin{figure}
  \centering
  \includegraphics[width=\columnwidth]{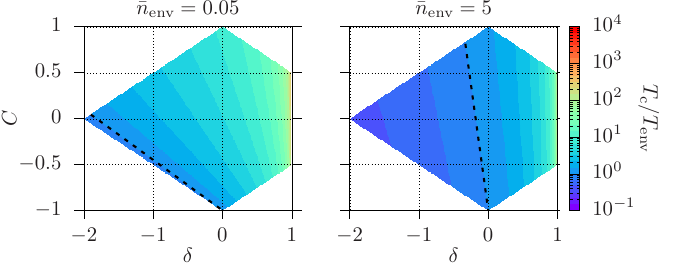}
  \caption{Steady-state cavity temperature $\Tc/\Tenv$ [Eq.~\eqref{eq_Tc}] as a function of the two-excitation inversion $\delta$ and the coherence $C$ for two different environmental temperatures corresponding to $\nenv=0.05$ (left) and $\nenv=5$ (right) thermal photons, respectively. The dotted black lines separate the cooling (left of the line) from the heating (right of the line) regime [Eq.~\eqref{eq_cooling_heating}]. Parameters: $\kappa=\mu/2$.}\label{fig_map}
\end{figure}

\par

Figure~\ref{fig_map} shows the behaviour of the temperature ratio $\Tc/\Tenv$ as a function of $\delta$ and $C$ in high- and low-temperature environments, respectively. There we have used $\kappa=\mu/2$. For $\mu\sim 10^6\,\text{s}^{-1}$ in the optical domain~\cite{quan2006quantum} this value corresponds to cavity loss rates or linewidths in the MHz regime. Depending on the two-atom coherence $C$, the cavity field may either be heated above the ambient (environment) temperature or cooled below the latter.

\par

It is further seen from Fig.~\ref{fig_map} that the heating regime, defined via the inequality $\Tc>\Tenv$, is significantly enhanced in a cold environment with very few thermal photons. Then, from Eq.~\eqref{eq_Tc}, we can show that the heating regime persists for
\begin{equation}\label{eq_cooling_heating}
  C>-1-\left(\nenv+\frac{1}{2}\right)\delta.
\end{equation}
This inequality is one of the central results of this paper. It implies that the benefit of HECs is maximal in low-temperature environments. It also shows how $C$ and $\delta$ complement each other as heating resources. The map in Fig.~\ref{fig_map} reflects the possible benefits of two-atom coherence as a resource for engineering effective heat baths using non-equilibrium states. We note that although the coefficients~\eqref{eq_r_pm} do not explicitly depend on the populations $\rho_{22}$ and $\rho_{33}$ of the single-excitation subspace, their values nevertheless determine the maximally-allowed coherence $C$ via the condition $|C|\leq 2\sqrt{\rho_{22}\rho_{33}}$ for $\rho$ in Eq.~\eqref{eq_rho} to be non-negative. This condition gives rise to (see Appendix~\ref{app_C_delta})
\begin{equation}\label{eq_C_rho}
  |C|\leq 1-\frac{|\delta|}{2}.
\end{equation}

\par

We can now address the core issue: What role is played by the two-atom coherence $C$ (that may pertain to classical correlations or entanglement between the atoms) for a given dimer energy (determined by $\delta$) in driving the cavity field to a thermal steady state? As a simple example, we consider the two singly-excited Bell states as compared to their phase-averaged counterpart,
\begin{subequations}
  \begin{align}
    \ket{\Psi^\pm}&\coloneq \frac{\ket{ge}\pm\ket{eg}}{\sqrt{2}}\label{eq_psi_pm} \\
    \rho_\mathrm{mix}&\coloneq \frac{1}{2}\proj{ge}+\frac{1}{2}\proj{eg}.\label{eq_rho_mix}
  \end{align}
\end{subequations}
All these states have the same energy and yet they result in very different temperatures
\begin{equation}\label{eq_T_pm_mix}
  T_+>T_\mathrm{mix}>T_-\equiv\Tenv,
\end{equation}
where $T_\pm$ pertain to $\ket{\Psi^\pm}$ and $T_\mathrm{mix}$ to $\rho_\mathrm{mix}$. Relation~\eqref{eq_T_pm_mix}, which shows the ability of two-atom coherence to enhance heating, can be easily obtained from Eq.~\eqref{eq_Tc} by substituting $\delta=0$ and $C=\pm 1$ [for the states~\eqref{eq_psi_pm}] and $C=0$ [for the state~\eqref{eq_rho_mix}], respectively. The same inequality~\eqref{eq_T_pm_mix} can be inferred from Fig.~\ref{fig_map}, where the highest temperature for $\delta=0$ is achieved for the Bell state $\ket{\Psi^+}$ [Eq.~\eqref{eq_psi_pm}]. The difference between $T_+$, $T_\mathrm{mix}$ and $T_-$ increases the colder the environment, i.e, the smaller $\nenv$ is.

\par

The fact that the Bell state $\ket{\Psi^+}$ yields the highest possible temperature (among all states with $\delta=0$) whereas its counterpart $\ket{\Psi^-}$ cannot modify the cavity temperature beyond the environmental temperature is a consequence of the symmetries of the underlying resonant Jaynes--Cummings Hamiltonian~\cite{wallsbook} that describes the interaction of the atom beam with the single-mode cavity field: Since the individual atoms are assumed to be indistinguishable for the cavity field~\cite{dag2016multiatom}, quantum interference gives rise to dark and bright states as known from Dicke superradiance~\cite{dicke1954coherence,lehmberg1970radiation2,mandelbook}. Whereas the Bell state $\ket{\Psi^+}$ maximizes energy transfer from the atom pair to the cavity mode (for states with $\delta=0$), destructive interference prevents any such transfer for a pair in the $\ket{\Psi^-}$ state, such that for this state $r_\pm=0$ and hence $T_-\equiv \Tenv$.

\section{Naturally entangled dimers}\label{sec_dimers}

\subsection{Molecular dimers}

According to the theory of electronically-excited molecular-dimer (homonuclear-diatom) dissociation~\cite{kurizki1985quantum,kurizki1987theory} there is a simple correspondence between the symmetry of the dimer state~\cite{herzbergbook} and the entanglement of the atoms emerging from the dissociation~\cite{kurizki1985quantum,kurizki1987theory,fry1995proposal,urbanczyk2012entangled}. Entanglement can emerge in dissociation fragments (or collision products) receding beyond the molecular interaction region. Correlations are directly determined by the state of the initial molecular dimer. The correlations of two identical fragments carrying a single joint excitation are determined by the symmetry and the spin of the parent dimer and by the angular momentum states of the fragments. Depending on these properties, the dissociated dimer may emerge in super- or subradiant Dicke states~\cite{kurizki1987theory}. This simple description mainly applies to systems like $\mathrm{Ca}_2$ that dissociate via single-excitation states $^1\Sigma_{u(g)}$ or $^1\Pi_{u(g)}$. The ungerade $(u)$ or gerade ($g$) symmetry then results in a symmetric or antisymmetric combination of the two-atom states $\ket{^1P_m}_{1(2)}\ket{^1S}_{2(1)}$. Such systems may thus be described as two identical TLAs, owing to the lack of mixing of degenerate $^1P_m$ magnetic levels at large distances~\cite{kurizki1988cooperative}.

\par

We may summarize the correspondence between the ``parent'' (molecular-dimer) state and the ``nascent'' two-atom entanglement as follows:
\begin{itemize}
\item Type~1 entanglement: Gerade, spin-singlet dimer state (such as $^1\Sigma_g$ or $^1\Pi_g$) or ungerade, spin-triplet dimer state (such as $^3\Sigma_u$ or $^3\Pi_u$) $\Leftrightarrow$ dark atom-pair (Dicke-singlet) state $\ket{\Psi^-}$.
\item Type~2 entanglement: Ungerade, spin-singlet dimer state (such as $^1\Sigma_u$ or $^1\Pi_u$) or gerade, spin-triplet dimer state (such as $^3\Sigma_g$ or $^3\Pi_g$) $\Leftrightarrow$ bright atom-pair (Dicke-triplet) state $\ket{\Psi^+}$.
\end{itemize}
The two-sided arrow indicates that $\ket{\Psi^-}$ or $\ket{\Psi^+}$ may be prepared by the appropriate dimer-state dissociation (as demonstrated experimentally in Ref.~\cite{grangier1985quantum}) or conversely, they may be recombined by a collision to form those dimer states~\cite{deb1999formation}.

\par

\begin{figure}
  \centering
  \includegraphics[width=0.95\columnwidth]{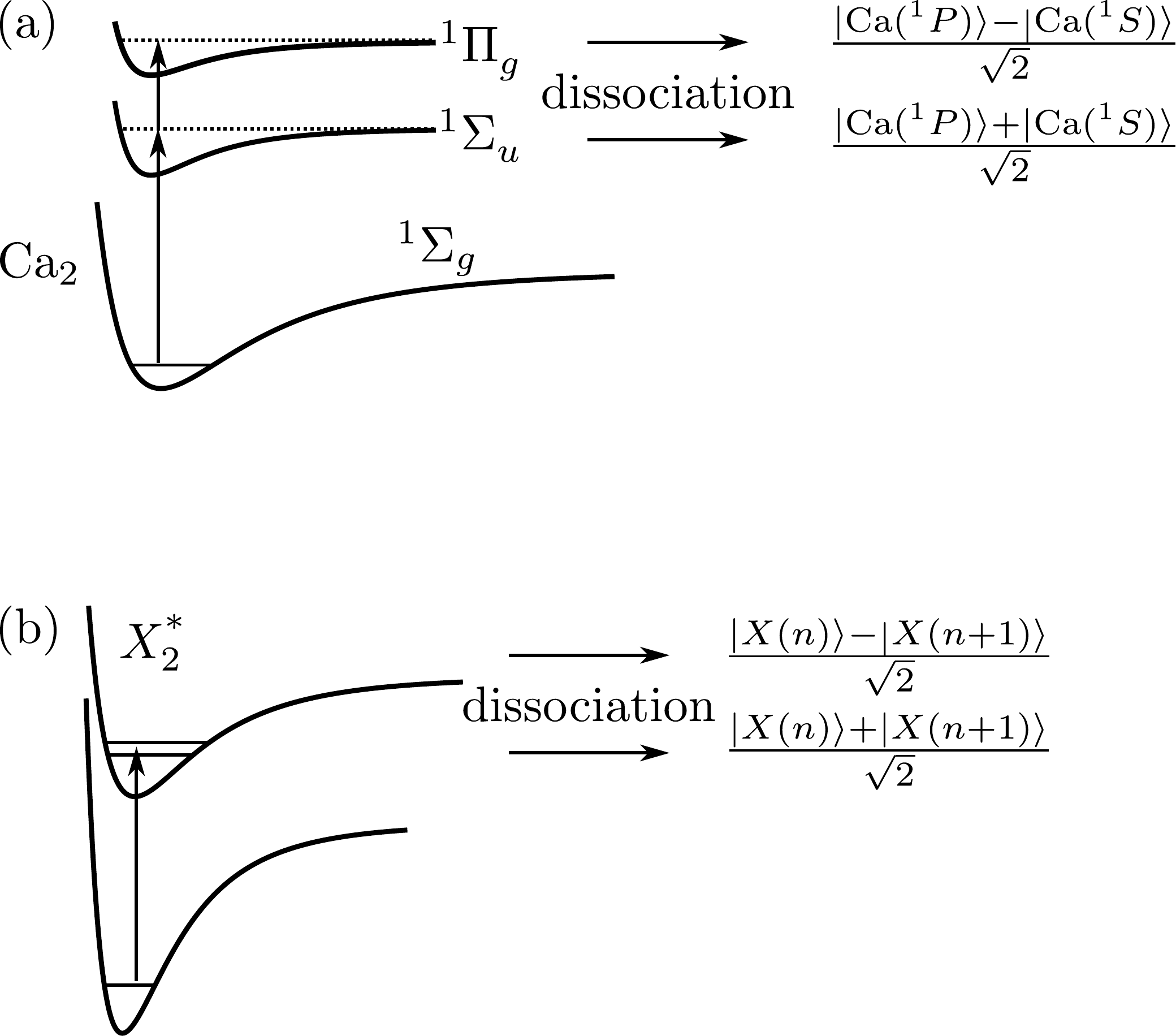}
  \caption{(a)~Preparation of a superradiant ($^1\Sigma_u$) or subradiant ($^1\Pi_g$) Dicke state by one- or two-photon excitation of $\mathrm{Ca}_2$ above the dissociation threshold. (b)~Idem for Rydberg states ($n\gg 1$) of any dimer $X_2$ that correlate to super- or subradiant Dicke states.}\label{fig_dimer}
\end{figure}

\par

The preparation protocol of these (type~1 or type~2) entangled atom-pair states is simple: Type~1 states require a two-photon (Raman) excitation of the dimer above dissociation threshold and type~2 states require one-photon (direct) excitation of the dimer above dissociation threshold (Fig.~\ref{fig_dimer}a).

\par

For resonant atom--field interaction in optical cavities the excited dimer should dissociate in two TLAs sharing an optical excitation, e.g., $\ket{\mathrm{Ca}_2^\text{*}(^1\Sigma_u)}\rightarrow (\ket{\mathrm{Ca}(^1P)}+\ket{\mathrm{Ca}(^1S)}/\sqrt{2}$ or $\ket{\mathrm{Ca}_2^\text{*}(^1\Pi_g)}\rightarrow (\ket{\mathrm{Ca}(^1P)}-\ket{\mathrm{Ca}(^1S)}/\sqrt{2}$ (Fig.~\ref{fig_dimer}a)~\cite{grangier1985quantum}.

\par

By contrast, in microwave cavities, such interactions require dimers that share a microwave excitation between Rydberg states, e.g., $\ket{X_2^\text{*}}\rightarrow (\ket{X(n)}\pm \ket{X(n+1)})/\sqrt{2}$ (Fig.~\ref{fig_dimer}b). In either case, the super- and subradiant Dicke states are predominantly split in energy by the two-atom resonant dipole--dipole interaction~\cite{lehmberg1970radiation2,petrosyan2002scalable}. At sufficiently large separations (exceeding sub-micron distance for optical excitations) the two emerging atoms do not interact and the energies of the different Dicke states become equal.

\par

The photoexcitation and dissociation may occur before the emerging atoms enter the cavity where they act as a bath on the cavity field [see Fig.~\ref{fig_hec_scheme} and Eq.~\eqref{eq_master}]. In what follows we outline a possible experimental implementation of the protocol described above. Molecular dimer dissociation can be effected by various mechanisms, such as beam-foil scattering or electron-impact excitation, but photodissociation by single-photon or two-photon (Raman) excitation is the most controllable means of attaining an electronically-excited dissociative state with a prescribed (odd or even, or equivalently, gerade or ungerade) exchange symmetry of the fragments (see Ref.~\cite{kurizki1987theory} and references therein). Following the photodissociation, the fragments (that share an electronic exciatation) go through a molecular interaction regime dominated by short-range interactions, followed by radiative intermediate and long-distance interaction regimes. Our model assumes that any cooperative effect due to either short-range or radiative coupling of the fragments is negligible. These assumptions are justified if the fragments arrive at the cavity after the dissociation when their distance $R$ to each other is comparable to or larger than the emission wavelength $\lambda\sim 100\,\text{nm}$ for an optical cavity~\cite{kurizki1987theory}. The radiative coupling of the fragments would then be a dipolar interaction with a strength comparable to the radiative linewidth of a fragment, ca.\ 10 to 100 MHz. Let us take as the speed of the fragments~\cite{kurizki1987theory} $v\sim 1\,\text{km/s}$. Such fragments can traverse a 1-micron cavity waist within a time of $\tau\sim 1\,\text{ns}$~\cite{quan2006quantum}. There can be only one pair of fragments, belonging to the same parent molecule, in the optical cavity if the repetition rate $r$ of the photodissociating (laser) pulses satisfies $r\leq 1/\tau \sim 1\,\text{GHz}$. Then, a cavity-dipole coupling strength of $g\sim 30\,\text{MHz}$~\cite{quan2006quantum} corresponds to $\mu=r(g\tau)^2\sim 1\,\text{MHz}$ as the effective coupling parameter in our theory when applied to optical resonators, same as the value used in the previous section (cf.\ Fig.~\ref{fig_map}). Taking $r\sim 1\,\text{GHz}$ gives a mean free time between the arrival of fragment pairs to the cavity of $1\,\text{ns}$. If the dissociation is effected by a $1\,\text{ns}$ pulse next to the cavity, the fragments will be separated by a distance $R\sim 1000\,\text{nm}$ when they arrive at the cavity after the dissociation, hence our model is applicable and the cooperative effects only stem from the initial state of the fragments. Inevitable delay of the arrival of the fragments to the cavity raises the issue of maintaining initial quantum entanglement since the excited fragments undergo radiative decay, and may also be exposed to a decohering environment (e.g., collisions) outside the cavity. Accordingly, it is necessary for us to address the question of quantum decoherence of entangled fragments during their transfer to the cavity from the dissociation site, which might take a few ns, shorter than a typical optical fluorescence lifetime of $\sim 10\,\text{ns}$. To this end, we will examine the effects of typical quantum decoherence channels on the entanglement dynamics in the next section. The main goal will be to determine how much time one needs to transfer sufficient entanglement to the cavity so that the cavity can still be ``heated'' by it (cf.\ Fig.~\ref{fig_temperature_decay}).

\subsection{Positronium dimers}

An analogous naturally-entangled system can be positronium~\cite{rich1981recent} obtained from the unstable bound state of an electron--positron pair. It can be produced experimentally~\cite{rich1981recent} in parapositronium and orthopositronium states, which are the corresponding Bell-type entangled singlet and triplet states~\cite{acin2001three,harpen2003positronium}. Their random production is compatible with the requirements of the micromaser theory of gamma-ray lasers~\cite{mills2004prospects}. The challenge of constructing a cavity for gamma photons that would serve as the working fluid of a gamma micromaser has been addressed within X-ray research~\cite{thomas2014optical}. Positronium clusters as sources for superradiance have also been considered~\cite{cui2012manipulating}.

\section{Dimer decoherence}\label{sec_decoherence}

We now examine the impact of quantum decoherence channels~\cite{nielsenbook} that act on the dimer beam on the achievable final cavity temperature~\eqref{eq_Tc}. More precisely, we consider phase-damping channels (PDCs), giving rise to dephasing, and generalized amplitude damping channel (GADCs), which are a combination of amplitude amplifying and amplitude damping channels. Physically, such a channel may correspond to the interaction of the atom pairs with the surrounding electromagnetic field at temperature $\Tenv$~\cite{wallsbook}. Note that although the matrix elements change, the structure of $\rho$ as specified in Eq.~\eqref{eq_rho} is preserved under both PDC and GADC.

\par

\begin{figure}
  \centering
  \includegraphics[width=\columnwidth]{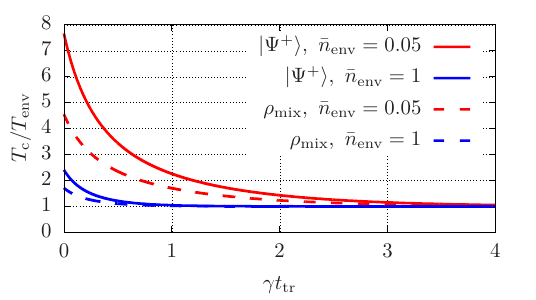}
  \caption{Achievable cavity temperature under the influence of dissipation on the atom beam during the transfer time $\ttr$ to the cavity. On their way to the cavity, the atoms interact with the surrounding electromagnetic field, resulting in emission and absorption of photons, modifying the dimer parameter according to Eqs.~\eqref{eq_C_delta_decay}. The initial state of the dimers is $\ket{\Psi^+}$ (Eq.~\eqref{eq_psi_pm}; solid lines) and $\rho_\mathrm{mix}$ (Eq.~\eqref{eq_rho_mix}; dashed lines), respectively. The red (upper two) lines correspond to a low-temperature environment with $\nenv=0.05$ photons and the blue (lower two) lines depict the situation for $\nenv=1$. The cavity temperature $\Tc$ is scaled by the respective environmental temperate $\Tenv$ and $\kappa=\mu/2$.}\label{fig_temperature_decay}
\end{figure}

\par

Under the influence of a PDC (e.g., non-radiative dephasing of the atoms due to collisions~\cite{wallsbook}), the initial coherence $\rho_{23}(0)$ decays exponentially, $\rho_{23}(\ttr)=\rho_{23}(0)\exp(-8\gamma_\mathrm{d} \ttr)$~\cite{wallsbook}, where $\ttr$ is the transfer time of the atom pair from their creation point to the cavity (assumed to be much longer than the interaction time $\ttr\gg\tau$) and $\gamma_\mathrm{d}$ the single-atom dephasing rate (see Appendix~\ref{app_dephasing}). Consequently, starting from the state $\ket{\Psi^+}$ [Eq.~\eqref{eq_psi_pm}] the achievable cavity temperature~\eqref{eq_Tc} decreases the more mixed the state becomes, i.e., the more $C(\ttr)=2\realp\rho_{23}(\ttr)$ is reduced. Hence, whilst PDCs do not alter $\delta$, their effect on $C$ limits the achievable temperatures beyond the ideal case depicted in Fig.~\ref{fig_map}.

\par

Let us now consider the interaction of the atomic pairs with the surrounding electromagnetic field at temperature $\Tenv$, resulting in incoherent absorption and emission processes~\cite{wallsbook}, which constitutes a GADC. In contrast to pure phase decay, now both $\delta$ and $C$ are altered due to decoherence. Denoting the spontaneous emission rate of a single atomic constituent by $\gamma$, these parameters evolve according to (see Appendix~\ref{app_decay})
\begin{subequations}\label{eq_C_delta_decay}
  \begin{align}
    C(\ttr)&=C(0)e^{-\gamma(2\nenv+1)\ttr}\\
    \delta(\ttr)&=\left[\frac{2}{2\nenv+1}+\delta(0)\right]e^{-\gamma(2\nenv+1)\ttr}-\frac{2}{2\nenv+1}.
  \end{align}
\end{subequations}
Hence, while $|C(\ttr)|$ again exponentially decreases with $\ttr$, now the inversion (or ``polarization'') $\delta(\ttr)$ also relaxes to a value determined by the environmental excitation $\nenv$. Figure~\ref{fig_temperature_decay} shows the ratio $\Tc(\ttr)/\Tenv$ as a function of $\gamma\ttr$ for atoms initialized in the Bell state $\ket{\Psi^+}$ [Eq.~\eqref{eq_psi_pm}] and in the phase-averaged state $\rho_\mathrm{mix}$ [Eq.~\eqref{eq_rho_mix}], respectively.

Generally, PDCs are less limiting than GADCs as they only affect $C$ but leave $\delta$ untouched. Consequently, under the action of PDCs, the final temperature tends towards the temperature induced by the phase-averaged state, which may still be higher than the environment temperature $\Tenv$ as in Eq.~\eqref{eq_T_pm_mix} for the states carrying a single excitation. By contrast, under the action of GADCs the temperature tends towards $\Tenv$. In any case, in order to avoid decoherence effects on the beam the transfer time of the atom pairs to the cavity should be faster than the inverse decoherence rate, $t_\mathrm{tr}\ll 1/\gamma$. This requirement can be easily satisfied for a dilute beam of dimers in which unwarranted collision effects are rare, and the beam velocity is supersonic~\cite{amirav1979intramolecular}.

\section{Comparison to phaseonium}\label{sec_comparison_phaseonium}

\begin{figure}
  \centering
  \includegraphics[width=0.95\columnwidth]{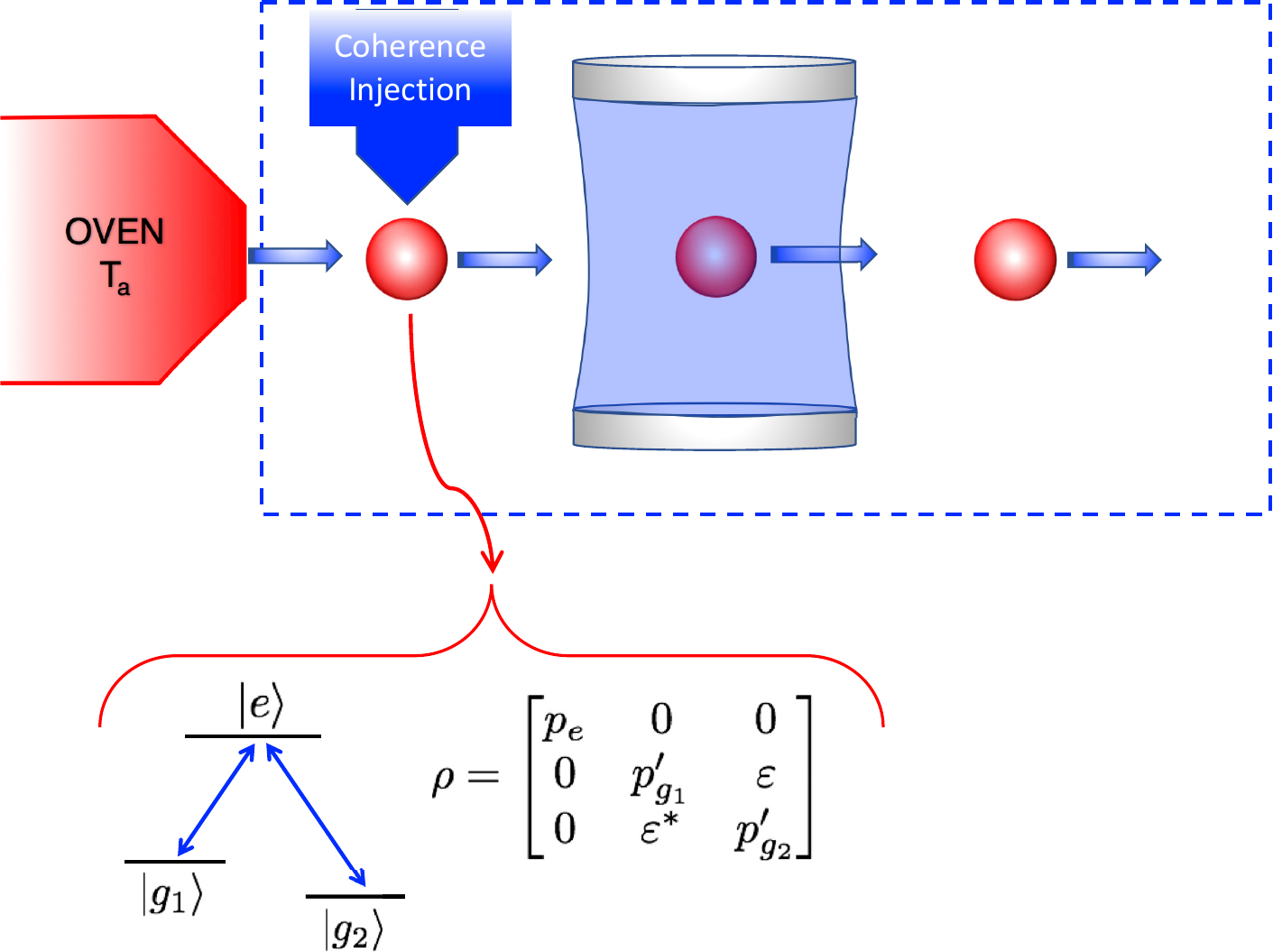}
  \caption{A micromaser powered by phaseonium quantum fuel. The setup consists of a cavity pumped by a beam of thermal three-level atoms at temperature $\Ta$ (with populations $p_e$, $p_{g_1}$ and $p_{g_2}$) to which some coherence $\varepsilon$ is injected before their interaction with the cavity mode. During the coherence injection the ground-state populations are slightly changed to $p_{g_1}^\prime$ and $p_{g_2}^\prime$, respectively.}\label{fig_phaseonium}
\end{figure}

In their seminal paper~\cite{scully2003extracting}, Scully et al.\ considered a micromaser setup where a beam of three-level atoms with two almost-degenerate ground states traverses the cavity. The atoms are initially in a thermal state at temperature $\Ta$ before a small amount of coherence is injected to the ground-state manifold (the resulting phase-coherent atoms have been nicknamed ``phaseonium''~\cite{scully2003extracting}), as depicted in Fig.~\ref{fig_phaseonium}. The cavity field state $\rho$ then evolves according to the following master equation~\cite{scully2002extracting}
\begin{multline}\label{eq_master_phaseonium}
  \dot\rho=\left[\frac{\mu}{4}\left(p_{g_1}+p_{g_2}+2|\varepsilon|\cos\varphi\right)+\kappa(\nenv+1)\right]\mathcal{L}_\mathrm{d}\rho\\+\left[\frac{\mu}{2} p_e+\kappa\nenv\right]\mathcal{L}_\mathrm{e}\rho,
\end{multline}
where $\mathcal{L}_\mathrm{d}$ again describes the de-excitation of the cavity mode due to photon emission and $\mathcal{L}_\mathrm{d}$ the excitation of the cavity mode owing to photon absorption. Here $p_e$, $p_{g_1}$ and $p_{g_2}$ denote the (thermal) excited and ground states populations of the three-level atoms and $\varepsilon=|\varepsilon| e^{i\varphi}$ is the coherence between the two ground states (see Fig.~\ref{fig_phaseonium}). Note that $p_{g_1}=\exp[-\hbar\Delta_g/(\kB\Ta)]p_{g_2}$, where $\hbar\Delta_g\coloneq\hbar\omega_{g_1}-\hbar\omega_{g_2}>0$ is the (small) energy splitting of the ground-state manifold.

\par

As in the conventional micromaser~\cite{filipowicz1986theory,scullybook,meystrebook}, the cooling and heating rates are modified by the respective atomic populations. Here, however, the cooling rate is additionally modified by the coherence and may be increased or decreased by adjusting the phase $\varphi$~\cite{scully2003extracting}. By contrast, the heating rate is not altered by the coherence, which is expected since the coherence pertains to the ground-state manifold. The highest achievable cavity temperature thus corresponds to the maximum reduction of the cooling rate, i.e., to the phase $\varphi=\pi$~\cite{scully2003extracting}.

\par

Phaseonium fuel hence increases the temperature of the cavity $\Tc$ beyond the atomic temperature $\Ta$ by reducing the cooling rate by $2|\varepsilon|$ compared to incoherent thermal atoms. The magnitude of the coherence, however, cannot be chosen arbitrary and must fulfill (see Appendix~\ref{app_phaseonium})
\begin{equation}\label{eq_phaseonium_max_coherence}
  2|\varepsilon|<p_{g_2}-p_{g_1}=p_{g_2}(1-\exp[-\hbar\Delta_g/(\kB\Ta)]).
\end{equation}

\par

In addition to the master equation, the effect of the coherence can be seen explicitly in the evolution of $\langle n\rangle$, the mean number of photons in the cavity, which is governed by
\begin{multline}\label{eq_ndot_phaseonium}
  \frac{\dd}{\dd t}{\langle n\rangle}=\mu p_e+2\kappa\nenv\\-2\left(\kappa+\frac{\mu}{4}\left(p_{g_1}+p_{g_2}-2p_e\right)+\frac{\mu}{2}|\varepsilon|\cos\varphi\right)\langle n\rangle,
\end{multline}
which in steady state yields
\begin{equation}
  \langle n\rangle_\mathrm{ss}=\frac{2\kappa\nenv+\mu p_e}{2\kappa+\frac{\mu}{2}(p_{g_1}+p_{g_2}-2p_e)+\mu|\varepsilon|\cos\varphi}.
\end{equation}
The first line in Eq.~\eqref{eq_ndot_phaseonium} drives $\langle n\rangle$ and is independent of the quantum coherence. The second line has a contribution from quantum interference effect, which favors enhancement of $\langle n\rangle$ for negative coherences. Equation~\eqref{eq_ndot_phaseonium} implies that the micromaser threshold depends on the coherence via
\begin{equation}\label{eq_phaseonium_threshold}
  2\kappa+\frac{\mu}{2}(p_{g_1}+p_{g_2}-2p_e)+\mu|\varepsilon|\cos\varphi>0.
\end{equation}

\par

The master equation~\eqref{eq_master_phaseonium} must be contrasted with the master equation~\eqref{eq_master} for a beam of correlated atoms carrying HECs (Fig.~\ref{fig_hec_scheme}). Since the coherence in the state~\eqref{eq_rho} is between states that carry one excitation quanta $\hbar\omegac$, both the cooling and heating rates in Eq.~\eqref{eq_master} are modified by the coherence via the parameter $C$. For this reason one may refer to $\rho_{23}$ as a ``hot'' coherence in contrast to the ``cold'' coherence in phaseonium. The threshold condition $\delta<2\kappa/\mu$ does not depend on the coherence $C$, contrary to phaseonium where $|\varepsilon|\cos\varphi$ affects the threshold~\eqref{eq_phaseonium_threshold}

\par

The mean number of cavity photons evolves according to
\begin{equation}\label{eq_ndot_hec}
  \frac{\dd}{\dd t}{\langle n\rangle}=\left(2\kappa\nenv+\mu r_+\right)-\left(2\kappa-\mu\delta\right)\langle n\rangle,
\end{equation}
which yields the mean steady-state photon number
\begin{equation}\label{eq_nss_hec}
  \langle n\rangle_\mathrm{ss}=\frac{2\kappa\nenv+\mu r_+}{2\kappa-\mu\delta}.
\end{equation}
In contrast to Eq.~\eqref{eq_ndot_phaseonium}, now the coherences only contribute to the first term in Eq.~\eqref{eq_ndot_hec}, i.e., to the driving.

\par

\begin{figure}
  \centering
  \includegraphics[width=0.95\columnwidth]{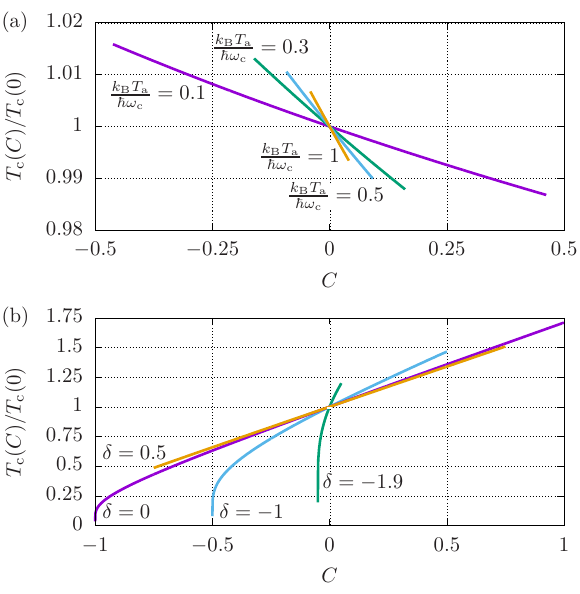}
  \caption{Cavity steady-state temperature for (a)~phaseonium and (b)~atom pairs as a function of the coherence. The temperature is scaled by the temperature without coherence. For phaseonium the values of $C=2|\varepsilon|\cos\varphi$ are mainly limited by the unitary injection process whereas the HEC of the dimer is limited by the non-negativity of $\rho$. Parameters: $\kB\Tenv=0.05\hbar\omegac$ (such that $\nenv\approx 0$), $\Delta_g=0.1\omegac$ and $\kappa=\mu/2$.}
  \label{fig_phaseonium_hec}
\end{figure}

\par

Figure~\ref{fig_phaseonium_hec} shows that the temperature ratio available by two-atom HECs is strongly increased compared to the phaseonium case where Eqs.~\eqref{eq_phaseonium_max_coherence} and~\eqref{eq_phaseonium_threshold} severely constrain the coherence $|\varepsilon|$. On the other hand, the preparation energy of a phaseonium atom is typically smaller than for a Bell state of two two-level atoms as the former corresponds to an energy splitting of $\hbar\Delta_g$ whereas the latter carries an excitation of frequency $\omegac$. One should however keep in mind that the respective preparation mechanisms strongly differ: Whereas the coherences in phaseonium need to be externally injected (e.g., by a microwave field~\cite{zubairy2002photo}), entangled atom pairs ``naturally'' emerge in dimer dissociation processes. Namely, the dissociating photon energy and the molecular symmetry imposed selection rules~\cite{kurizki1985quantum,kurizki1987theory,herzbergbook,fry1995proposal,urbanczyk2012entangled,grangier1985quantum} can unequivocally select the desired entangled state.

\par

Hence, while the heating ($\Tc>\Tenv$) regime~\eqref{eq_cooling_heating} does not necessarily require the atoms to be entangled, the most beneficial state for a given $\delta$ (i.e., for a given energy), namely, the state for which $\Tc(C)/\Tc(0)$ is maximized is the state with the maximally-possible coherence $C=1-|\delta|/2$ (cf.\ Eq.~\eqref{eq_ndot_hec} and Fig.~\ref{fig_phaseonium_hec}b) which is always entangled (see Appendix~\ref{app_heating_entanglement}).

\par

The possibility of using a beam of entangled atoms to control the cavity temperature has been previously proposed in Ref.~\cite{dillenschneider2009energetics}. However, contrary to our scheme where the atom pair is prepared in a general state of the form~\eqref{eq_rho}, in Ref.~\cite{dillenschneider2009energetics} thermal entangled states are considered. As shown above, coherences in the general initial state~\eqref{eq_rho} may lead to both cooling and heating effects (Fig.~\ref{fig_map}) that may be beneficial to engine operation, while thermal entangled pairs in Ref.~\cite{dillenschneider2009energetics} always result in cooling if the atoms are injected into the cavity as a pair.

\par

The use of phaseonium as a quantum fuel is severely restricted by atomic dephasing inside the cavity, which would limit the value of coherence to the range of $0.1$ for the typical resonators~\cite{quan2006quantum}. In contrast, as shown in Sec.~\ref{sec_decoherence}, dimers can still serve as beneficial fuel with HECs in this dephasing-limited regime. If the atomic dephasing limits $C$ to $\sim 0.1$, we can still get $\sim10\%$ enhancement of the cavity temperature $T_\mathrm{c}$ according to Fig.~\ref{fig_phaseonium_hec}.

\section{Discussion and conclusions}\label{sec_discussion}

We have investigated a dissipative micromaser driven by a beam of either classically-correlated or entangled atom pairs (dimers) subject to decoherence. Controlled by the two-atom coherence, the leaky cavity may be strongly heated above or cooled below the temperature of the environment. The wide temperature range resulting from such control (Fig.~\ref{fig_map}) suggests that entangled atom pairs may be exploited as an effective high-temperature or highly caloric fuel for quantum heat engines. In particular, we find that the strongest effect of two-atom coherence pertains to the case of atoms entangled in the Bell state $\ket{\Psi^+}$ (Fig.~\ref{fig_phaseonium_hec}b) that are unequivocally obtainable by dissociation of electronically-excited molecular dimers (e.g., $\mathrm{Ca}_2$) as in Fig.~\ref{fig_dimer}a or by the decay of the unstable bound states of positronium~\cite{rich1981recent,acin2001three,harpen2003positronium,mills2004prospects,thomas2014optical}. The resilience of the beam to atomic decoherence via phase-damping channels (PDCs) or generalized amplitude-damping channels (GADCs) has been found to be high, so that coherence/entanglement effects may prevail for sufficiently fast transfer times to the cavity (Fig.~\ref{fig_temperature_decay}).

\par

We have compared the micromaser fuelled by correlated dimers with its counterpart fuelled by phaseonium comprised of coherent three-level atoms~\cite{scully2003extracting}. In general, we find that the achievable cavity temperature range is larger for the proposed dimer-based micromaser (Fig.~\ref{fig_phaseonium_hec}), much because dimer coherence is free of the inherent constraints that phaseonium coherence is subject to (compare Eqs.~\eqref{eq_cooling_heating} and~\eqref{eq_C_rho} for the HEC with Eqs.~\eqref{eq_phaseonium_max_coherence} and~\eqref{eq_phaseonium_threshold} for phaseonium). Coherent dimers are also more ``natural'' to control and to prepare than phaseonium, as shown here. In the case of phaseonium, increasing the coherence between the lower levels would require weakly excited atoms, which would make the atom incapable of carrying much energy into the cavity.

\par

Heat-exchange coherences (HECs) in entangled dimers may find unique applications in quantum technologies: 

(a)~\emph{Heat machines}: If the cavity mode is coupled to only one environment at temperature $\Tenv$, we cannot operate a heat engine, but an incoherent (mixed-state) beam of atom pairs may form a local bath for the cavity mode at a temperature $T_\mathrm{mix}>\Tenv$ [Eq.~\eqref{eq_T_pm_mix}], so that this mode my alternately interact with baths at $\Tenv$ and $T_\mathrm{mix}$. The Carnot bound~\cite{schwablbook} on the heat-engine efficiency is then $\eta_\mathrm{C}^\mathrm{mix}=1-\Tenv/T_\mathrm{mix}$. By contrast, a beam of entangled dimers in the Bell state $\ket{\psi^+}$ [Eq.~\eqref{eq_psi_pm}] may raise the Carnot bound to $\eta_\mathrm{C}^+=1-\Tenv/T_+$. If, for example, $\nenv=0.6$ such that $\eta_\mathrm{C}^\mathrm{mix}\approx 1/2$, we can boost it by $33\%$ to $\eta_\mathrm{C}^+\approx 2/3$. 
In the proposed dimer dissociation scheme (Fig.~\ref{fig_dimer}) the excitation energy of the photons that cause dissociation exceeds the cavity-photon energy $\hbar\omegac$. However, following the thermodynamic tradition~\cite{cengelbook}, these bath preparation costs are not included in the above efficiency considerations~\cite{scully2003extracting,dillenschneider2009energetics,abah2014efficiency,
rossnagel2014nanoscale,turkpence2016quantum,niedenzu2016operation,niedenzu2018quantum,klaers2017squeezed}.
In the long run, \emph{positronium-based gamma-ray micromasers} (Sec.~\ref{sec_dimers}) are an enticing prospect. 

\par

(b)~\emph{Quantum annealers}: HEC in dimers may provide fast and broadly-tunable baths for quantum annealing~\cite{amin2008thermally}.
While it is in principle suitable for low-temperature implementations, such as superconducting qubits, it may be more resilient to errors at higher temperatures~\cite{kechedzhi2016open,nishimura2016retrieving}. High-temperature quantum annealers would benefit from artificial quantum thermal baths with widely tunable temperature~\cite{shabani2016artificial}. Natural thermal reservoirs lead to slow thermalization times, determined by non-unitary open-system dynamics; hence, unitary generation and resetting of effective thermal baths are highly desired~\cite{shabani2016artificial}.
More specifically, it is proposed in Ref.~\cite{shabani2016artificial} to use superconducting resonators as thermal baths for finite-temperature quantum annealers. It is, however, also pointed out that resetting the resonator via incoherent thermalization takes too long and hence is the bottleneck against high-speed thermal quantum annealing. We here use a similar physical system based on a cavity, but instead of thermalizing it by incoherent drives we consider quantum dimers that coherently interact with the cavity mode at random times. Hence, our scheme can be used for the coherent preparation of thermal states of the resonator, which can serve as a bath for thermal quantum annealers. Accordingly, our scheme can therefore be a potential solution for a fast coherent route to thermalization in finite-temperature quantum annealing (see Appendix~\ref{app:thermalTime}).

\par
 
(c)~\emph{Quantum simulators of photosynthesis}: One may simulate the dependence of energy exchange on the bath spectrum and temperature by controlling HECs in the dimers. In general, while here we have restricted ourselves to naturally-entangled dimers, an extension to highly-controllable synthetic quantum dimers, such as entangled transmon qubits, as heat sources for quantum devices would be interesting.
A recent experiment has realized a quantum simulator for light harvesting complexes with superconducting circuits~\cite{potocnik2018studying}. It has been argued that the non-locality of quantum entanglement or coherence may play a role in the fast energy transfer through the complex~\cite{huelga2013vibrations}. Dimer fuels, as proposed here, can serve as an effective bath in such quantum biological simulators. Under noisy dissipative conditions such dimers can mimic a background molecular system whose heat acts as a spectrally structured reservoir on the energy transfer.

\par

The overarching goal of our consideration of quantum fuels is to develop a framework for a genuine quantum thermodynamical approach to quantum devices, in contrast to semi-classical thermodynamical machines wherein classical resources drive quantum systems, which can be compared to the difference between semi-classical and genuine quantum optics.

\par

\begin{acknowledgments}

F.\,O.\ and \"{O}.\,E.\,M.\ acknowledge support by TUBITAK (Grant No.\ 116F303) and by the EU-COST Action (CA15220). F.\,O.\ acknowledges the Isik University Scientific Research Fund (Grant No.\ BAP-15B103) for support. W.\,N.\ acknowledges support from an ESQ fellowship of the Austrian Academy of Sciences (\"OAW). G.\,K. acknowledges support by the ISF and the DFG support through the project FOR 2724.

\end{acknowledgments}

\appendix

\section{Derivation of the master equation}\label{app:derivation}

Under the standard assumptions of micromaser theory~\cite{meystrebook,scullybook}, master equations describing the cavity field dynamics pumped by two- and three-atom clusters have been derived in Ref.~\cite{dag2016multiatom}. Here we outline the derivation of the master equation for the dimer case in the presence of cavity loss.

\par

We consider the Tavis-Cummings model~\cite{tavis1968exact} $H_{\mathrm{TC}} = H_\mathrm{a} + H_\mathrm{c} + H_{\text{int}}$ with the atom, cavity and interaction Hamiltonians
\begin{subequations}
  \begin{align}
    H_\mathrm{a} & = \hbar \omega_\mathrm{a} \sum_{k=1}^2 \sigma_k^+\sigma_k^- \label{eq_app_Ha}\\
    H_\mathrm{c} &= \hbar \omega_\mathrm{c} a^{\dagger} a \\
    H_{\text{int}} &= \hbar g \sum_{k=1}^2 ( a \sigma_k^+ + a^{\dagger} \sigma_k^-).
  \end{align}
\end{subequations}
Here $a,a^{\dagger}$ are the annihilation and creation operators for the cavity field and $\sigma^+_k, \sigma^-_k$ are the raising and lowering Pauli operators for the $k$th atom ($k=1,2$), respectively. We consider only resonant operations so that $\omega_\mathrm{a}=\omega_\mathrm{c}$. The interaction strength $g$ between the atoms and the cavity is assumed to be homogeneous.

\par

During the short interaction time $\tau$ the whole system evolves unitarily. The propagator $U(\tau)$ for the Hamiltonian $H_{\text{int}}$ can be analytically determined~\cite{dag2016multiatom}. Denoting the arrival time of the $j$th dimer to the cavity by $t_j$, the cavity-field density matrix $\rho_\mathrm{c}$ evolves according to~\cite{gardinerbook}
\begin{equation}
  \rho_\mathrm{c} (t_j + \tau) = \text{Tr}_\mathrm{a}\left[U(\tau) \rho \otimes \rho_\mathrm{c} (t_j) U^{\dagger}(\tau)\right]=: S(\tau) \rho_\mathrm{c}(t_j).
\end{equation}
Here $\rho$ is the initial density operator of the dimers and $S(\tau)$ is a superoperator. The dimers pass through the cavity within a time interval of $(t,t+\delta t)$ with a probability of $p \delta t$. When a dimer passes through the cavity, the cavity field evolves according to $S(\tau)$ such that
\begin{equation}
  \rho_\mathrm{c}(t+\delta t) = p \delta t S(\tau) \rho_\mathrm{c}(t) + (1-p\delta t) \rho_\mathrm{c}(t).
\end{equation} 
In the limit $\delta t \rightarrow 0$ we obtain the master equation~\cite{liao2010single,li2014quantum,quan2006quantum,filipowicz1986theory}
\begin{equation}\label{eq_app_master0}
  \dot{\rho}_\mathrm{c}(t) = p \left[ S(\tau) - 1 \right] \rho_\mathrm{c}(t).
\end{equation}

\par

In terms of the dimer initial state $\rho$ [Eq.~\eqref{eq_rho}], the master equation~\eqref{eq_app_master0} evaluates to~\cite{dag2016multiatom}
\begin{equation}
\dot{\rho}_\mathrm{c}(t) = p \left[ \sum_{i,j=1}^4 \rho_{ij} \sum_{n=1}^4 U_{ni}(\tau) \rho_\mathrm{c}(t) U_{nj}^{\dagger}(\tau)- \rho_\mathrm{c}(t) \right],
\end{equation}
which, using the explicit form of the evolution operator $U(\tau)$, yields the Lindblad master equation
\begin{equation}\label{eq_app_master1}
  \dot{\rho}_\mathrm{c} \approx \mu\left(\frac{r_+}{2} \mathcal{L}_\mathrm{e}\rho_\mathrm{c} + \frac{r_-}{2} \mathcal{L}_\mathrm{d}\rho_\mathrm{c} \right),
\end{equation}
where the superoperators $\mathcal{L}_\mathrm{d}\rho_\mathrm{c}$ and $\mathcal{L}_\mathrm{e}\rho_\mathrm{c}$ are defined below Eq.~\eqref{eq_master}. In terms of the elements of the density matrix~\eqref{eq_rho}, the coefficients $r_\pm$ read~\cite{dag2016multiatom}
\begin{subequations}
  \begin{align}
    r_+ &=2\rho_{11}+\rho_{22}+\rho_{33}+\rho_{23}+\rho_{32} \\
    r_- &=2\rho_{44}+\rho_{22}+\rho_{33}+\rho_{23}+\rho_{32}.
  \end{align}
\end{subequations}
In Eq.~\eqref{eq_r_pm} we further parameterize these coefficients in terms of the two-atom coherence $C$ and the double-excitation population inversion $\delta$. The reason is that in this way it is clearer to see the dependence of the temperature control on the quantum coherence and the energy of the dimers.

\par

In order to account for the coupling of the cavity-field mode to the environment we add the Liouvillian~\cite{wallsbook,gardinerbook}
\begin{equation}
  \mathcal{L}_\mathrm{c}\rho_\mathrm{c} = \kappa \left(\bar{n}_{\text{env}} + 1\right) \mathcal{L}_\mathrm{d}\rho_\mathrm{c} + \kappa \bar{n}_{\text{env}}\mathcal{L}_\mathrm{e}\rho_\mathrm{c}
\end{equation}
to the master equation~\eqref{eq_app_master1}, which now reads
\begin{equation}
  \dot{\rho}_\mathrm{c} = \mu\left(\frac{r_+}{2} \mathcal{L}_\mathrm{e}\rho_\mathrm{c} + \frac{r_-}{2} \mathcal{L}_\mathrm{d}\rho_\mathrm{c} \right)+\mathcal{L}_\mathrm{c}\rho_\mathrm{c}.
\end{equation}
This is the master equation~\eqref{eq_master} in the main text.

\section{Energy of the dimer}\label{app:dimerEnergy}

The energy of the dimer reads
\begin{equation}
  E=\Tr[H_\mathrm{a}\rho],
\end{equation}
where $H_\mathrm{a}$ is given in Eq.~\eqref{eq_app_Ha}. For the dimer state~\eqref{eq_rho}, assuming resonance, $\omega_\mathrm{a}=\omega_\mathrm{c}$, the energy evaluates to 
\begin{equation}
  E=\hbar\omega_\mathrm{c} \left( 2\rho_{11} + \rho_{22} + \rho_{33} \right)\equiv\hbar\omega_\mathrm{c}(1+\delta/2).
\end{equation}

\section{Maximal coherence}\label{app_C_delta}

The non-negativity of $\rho$ in Eq.~\eqref{eq_rho} requires $\rho_{23}^2\leq\rho_{22}\rho_{33}$. Parameterizing  $\rho_{33}=\alpha\rho_{22}$ with $\alpha\in(0,\infty)$, this condition yields $\rho_{22}\geq|\rho_{23}|/\sqrt{\alpha}$. Hence, $\rho_{22}+\rho_{33}=\rho_{22}(1+\alpha)\geq |\rho_{23}|(1+\alpha)/\sqrt{\alpha}\geq 2|\rho_{23}|=|C|$ since $(1+\alpha)/\sqrt{\alpha}$ has a minimum at $\alpha = 2$.

\par

The normalization of $\rho$ yields $1-\delta/2-\rho_{22}-\rho_{33}=2\rho_{44}\geq 0$. Hence, $1-\delta/2\geq\rho_{22}+\rho_{33}$, from which follows the condition
\begin{equation}\label{eq_app_condition_1}
  |C|\leq 1-\frac{\delta}{2}.
\end{equation}
On the other hand, the normalization of $\rho$ also yields $1+\delta/2-\rho_{22}-\rho_{33}=2\rho_{11}\geq 0$, from which follows the condition
\begin{equation}\label{eq_app_condition_2}
  |C|\leq 1+\frac{\delta}{2}.
\end{equation}
Combining Eqs.~\eqref{eq_app_condition_1} and~\eqref{eq_app_condition_2} then yields the bound
\begin{equation}
  |C|\leq 1-\frac{|\delta|}{2}.
\end{equation}

\section{Master equation for atomic dephasing}\label{app_dephasing}

The master equation for phase decay in two-level atoms reads~\cite{wallsbook}
\begin{equation}
  \dot{\rho}=\sum_{i=1}^2\gamma_\mathrm{d}\left(2\sigma_z^i\rho\sigma_z^i-2\rho\right),
\end{equation}
yielding
\begin{equation}
  \dot\rho_{23}=-8\gamma_\mathrm{d}\rho_{23}.
\end{equation}
Here $\gamma_\mathrm{d}$ is the single-atom dephasing rate induced by, e.g., atomic collisions~\cite{wallsbook}.

\section{Master equation for atomic decay}\label{app_decay}

The master equation for the independent decay of the two-level atoms in an environment at temperature $\Tenv$ reads~\cite{wallsbook}
\begin{multline}
  \dot{\rho}=\sum_{i=1}^2\frac{\gamma}{2}(\nenv+1)\left(2\sigma_-^i\rho\sigma_+^i-\sigma_+^i\sigma_-^i\rho-\rho\sigma_+^i\sigma_-^i\right)\\
  +\sum_{i=1}^2\frac{\gamma}{2}\nenv\left(2\sigma_+^i\rho\sigma_-^i-\sigma_-^i\sigma_+^i\rho-\rho\sigma_-^i\sigma_+^i\right),
\end{multline}
which yields the differential equations
\begin{subequations}
  \begin{align}
    \dot{C}&=-\gamma(2\nenv+1)C\\
    \dot{\delta}&=-\gamma(2\nenv+1)\delta-2\gamma,
  \end{align}
\end{subequations}
whose solution is given in Eqs.~\eqref{eq_C_delta_decay}. Here $\gamma$ denotes the single-atom spontaneous emission rate.

\section{Coherence in phaseonium}\label{app_phaseonium}

The coherence is unitarily injected into three-level atoms with thermal populations $p_e$, $p_{g_1}$ and $p_{g_2}$~\cite{scully2003extracting},
\begin{equation}
  \begin{pmatrix}
    p_e & 0 & 0 \\
    0 & p_{g_1} & 0 \\
    0 & 0 & p_{g_2}
  \end{pmatrix}
  \mapsto
  \begin{pmatrix}
    p_e & 0 & 0 \\
    0 & p_{g_1}-\xi & \varepsilon \\
    0 & \varepsilon^* & p_{g_2}+\xi
  \end{pmatrix}.
\end{equation}
The eigenvalues do not change under unitary transformations. The invariance of the determinant then yields the condition
\begin{equation}
  (p_{g_1}-\xi)(p_{g_2}+\xi)-|\varepsilon|^2=p_{g_1}p_{g_2},
\end{equation}
which further evaluates to
\begin{equation}
  \xi=\frac{p_{g_2}-p_{g_1}}{2}\pm\frac{1}{2}\sqrt{(p_{g_2}-p_{g_1})^2-4|\varepsilon|^2}.
\end{equation}
A real $\xi$ thus requires
\begin{equation}
  2|\varepsilon|<p_{g_2}-p_{g_1}=p_{g_2}(1-\exp[-\hbar\Delta_g/(\kB\Ta)]).
\end{equation}
This means that the higher $\Ta$ (i.e., the smaller $p_{g_2}-p_{g_1}$) the smaller the possibly injectable coherence $|\varepsilon|$. Note that $p_{g_2} = 1/Z$ where $Z = 1 + \exp \left[ -\delta_g/T_\mathrm{a} \right] +\exp \left[ -\omega_c/T_\mathrm{a} \right]$ is the partition function.

\section{Maximal heating and entanglement}\label{app_heating_entanglement}

According to Eq.~\eqref{eq_ndot_hec} and Fig.~\ref{fig_phaseonium_hec}b, the maximum heating effect of coherence pertains to the state~\eqref{eq_rho} with the maximally-possible $C$ for a given $\delta$ (see Appendix~\ref{app_C_delta}), namely
\begin{equation}
  \rho_\mathrm{h}=
  \begin{pmatrix}\label{eq_app_rhoh}
    \rho_{44}+\frac{\delta}{2} & 0 & 0 & 0 \\
    0 & \rho_{22} & \frac{1}{2}-\frac{|\delta|}{4} & 0 \\
    0 & \frac{1}{2}-\frac{|\delta|}{4} & \rho_{33} & 0 \\
    0 & 0 & 0 & \rho_{44}
  \end{pmatrix}
\end{equation}
for $-2\rho_{44}\leq\delta<2$ (since $\rho_{44}+\delta/2 \geq 0$ and $\delta<2$). This state is entangled~\cite{peres1996separability} iff $|\rho_{23}|^2 \geq \rho_{11} \rho_{44}$, i.e., iff $\frac{1}{2}-\frac{|\delta|}{4}>\sqrt{\rho_{44}\left(\rho_{44}+\frac{\delta}{2}\right)}$.

\par

From $\rho_{22}+\rho_{33}\geq 2 |\rho_{23}| = 1-\frac{|\delta|}{2}$ and the normalization of $\rho_\mathrm{h}$ follows
\begin{equation}
  2\rho_{44}+\frac{\delta}{2}=1-\rho_{22}-\rho_{33}\leq 1-\left(1-\frac{|\delta|}{2}\right)=\frac{|\delta|}{2},
\end{equation}
which yields the condition
\begin{equation}\label{eq_app_relation_rho_44}
  \rho_{44}\leq\frac{|\delta|-\delta}{4}.
\end{equation}

\par

Let us first consider the case $\delta\geq0$. Condition~\eqref{eq_app_relation_rho_44} then implies $\rho_{44}=0$ such that the state~\eqref{eq_app_rhoh} is always entangled for $0\leq \delta < 2$.

\par

In the case $\delta<0$, condition~\eqref{eq_app_relation_rho_44} implies $\rho_{44}\leq\frac{|\delta|}{2}$. The non-negativity of $\rho_\mathrm{h}$, however, requires $|\delta|\leq 2\rho_{44}$. Hence, for negative $\delta$ only $\rho_{44}=\frac{|\delta|}{2}$ is possible, implying entanglement of the state~\eqref{eq_app_rhoh}.

\par

The state~\eqref{eq_app_rhoh} is thus entangled for any $\delta\in[-2\rho_{44},2)$.

\section{Cavity thermalization timescale}\label{app:thermalTime}

Here we elaborate on the thermalization time-scale of the cavity. Let us choose an observable to be thermalized, e.g., the mean number of photons in the cavity, which evolves according to Eq.~\eqref{eq_ndot_hec}. Solving this equation for the initial vacuum state of the cavity, we obtain
\begin{equation}
  \langle n(t)\rangle = \langle n\rangle _{\text{ss}} \left[1-e^{-(2\kappa-\mu\delta)t} \right],
\end{equation}
where $\langle n\rangle _{\text{ss}}$, defined in Eq.~\eqref{eq_nss_hec}, is the steady-state photon number attained in the limit $t \rightarrow \infty$. While complete thermalization only occurs in this limit, practically, however, we can approximate the cavity thermalization time to be
\begin{equation}
  t_{\text{th}} \gg (2\kappa-\mu\delta)^{-1}.
\end{equation}
Hence, for negative $\delta$ the thermalization time may be strongly reduced compared to $2\kappa$.

\end{document}